\documentclass[twocolumn]{aastex62}
\usepackage{comment}
\usepackage{multirow}
\usepackage{booktabs}
\usepackage[flushleft]{threeparttable}
\usepackage[utf8]{inputenc} 
\usepackage{graphicx} 

\graphicspath{{./}{Figures/}}

\shorttitle{The Influence of Planet 9 on the Orbits of Distant TNOs}
\shortauthors{Cáceres \& Gomes}

\begin{document}

\title{The Influence of Planet 9 on the Orbits of Distant TNOs: The Case for a Low Perihelion Planet}

\correspondingauthor{Jessica Cáceres}
\email{jessica@on.br}

\author{Jessica Cáceres}
\affil{Observatório Nacional, Rua General José Cristino 77, CEP 20921-400 Rio de Janeiro, RJ, Brazil}

\author{Rodney Gomes}
\affiliation{Observatório Nacional, Rua General José Cristino 77, CEP 20921-400 Rio de Janeiro, RJ, Brazil}

\begin{abstract}

The hypothesis of an additional planet in the outer Solar System has gained new support as a result of the confinement noted in the angular orbital elements of distant trans-Neptunian objects. Orbital parameters proposed for the external perturber suggest semimajor axes between $500$ and $1000$ au, perihelion distances between $200$ and $400$ au for masses between $10$ and 20 $M_{\oplus}$. In this paper we study the possibility that lower perihelion distances for the additional planet can lead to angular confinements as observed in the population of objects with semimajor axes greater than 250 au and perihelion distances higher than 40 au. We performed numerical integrations of a set of particles subjected to the influence of the known planets and the putative perturber during the age of the Solar System and compared our outputs with the observed population through a statistical analysis. Our investigations showed that lower perihelion distances from the outer planet usually lead to more substantial confinements than higher ones, while retaining the Classical Kuiper Belt as well as the ratio of the number of detached with perihelion distances higher than $42$ au to scattering objects in the range of semimajor axes from $100$ au to $200$ au.

\end{abstract}


\section{Introduction}\label{sec:introduction}

The existence of trans-Neptunian objects (TNOs) with high perihelion distance ($q\gtrsim40$ au) and semimajor axis ($a\gtrsim200$ au) can not be explained by the sole influence of the known planets. Thus, several mechanisms have been proposed to explain the existence of these objects, whose best known member is Sedna ($a=482$ au and $q=76$ au) \citep{Brown2004}. Some of these mechanisms are: the interaction of the Sun with other stars when it was still in its birth cluster and stellar flybys were common (\cite{Brasser2012}, \cite{Dukes2012}), the capture of planetesimals from the external disk of another star (\cite{Kenyon2004}, \cite{Morbidelli2004}), a planet that could have existed temporarily in the scattered disk (\cite{Gladman2006}) or an existing yet undiscovered planet (\cite{Gomes2006}). 

From a few years ago several patterns have been noticed in distant TNOs supporting the hypotheses of an outer planet in the Solar System. After announcing the discovery of an object with perihelion distance similar to that of Sedna, 2012 VP$_{113}$ ($a=256$ au and $q=80$ au), \citet{Trujillo2014} noted that objects with $a>150$ au and $q>30$ au exhibit a confinement in their argument of perihelion ($\omega$). They argued that under the influence of the known planets the argument of perihelion of these objects would circulate in a fairly short and different from each other timescale. On the other hand, they showed that a super-Earth mass planet in a circular and low inclination orbit with semimajor axis between $200$ and $300$ au could lead to long period $\omega$- libration around $0^\circ$ for Sedna-like objects, but not for objects with lower perihelia. This opened the interesting possibility that other configurations of an outer planet might give the suitable tuning.

Later, \citet{Batygin2016} noted that TNOs with $a>250$ au and $q>30$ au are confined not only in $\omega$, but also in longitude of the ascending node ($\Omega$). They showed that the presence of a distant, eccentric, and massive planet could reproduce such confinements as well as perihelion detachment of distant TNOs as a consequence of secular and resonant interactions (see also \cite{Batygin2017} and references therein). This planet (hereafter named Planet 9, whose orbital parameters will be denoted with the subscript 9) would roughly share the same orbital plane as those of the confined objects and would be apsidally anti-aligned with them. Planet 9 would also predict the existence of apsidally aligned and high perihelion objects, if a wide distribution of perihelion is initially assumed (see also \cite{Khain2018}), and objects that would be compatible with the Centaurs of high semimajor axis and high inclination. Likewise, It offers an explanation for the origin of highly inclined TNOs ($i>60^\circ$) with $a<100$ au \citep{Batygin2016b} and the high inclination of the recently discovered long-period TNO, 2015 BP$_{519}$ ($q=36$ au) (\cite{Becker2018}, \cite{Batygin2017}). 

\cite{Brown2016} showed that confined TNOs orbits can be produced by a Planet 9 with the following orbital parameters\footnote{These numbers are taken from the paper's Figs (2) and (3) and their empirical expressions of $e_9$ as a function of $a_9$.}: 200 au $<q_{9}<$ $400$ au, 500 au $<a_{9}<$ $1000$ au, $i_{9}\sim30^\circ$, and mass $10 M_{\oplus}<m_{9}<20 M_{\oplus}$. By comparing the allowed orbital paths and estimated brightness of the planet to previous and ongoing surveys, they indicate that Planet 9 would be near aphelion with an apparent magnitude between $22$ and $25$. 

Subsequent works focused on restricting the location of Planet 9 on its orbit from the reduction of the residuals in Saturn's orbital motion based on Cassini data (\cite{Fienga2016}, \cite{Holman2016b}), or in Pluto's orbit (\cite{ Holman2016}), or assuming that several of the distant TNOs are in mean motion resonance (MMR) with the planet (\cite{Malhotra2016}, \cite{Millholland2017}).

An outer planet also provides a better explanation for the fact of there being an excess of bright large semimajor axis Centaurs with respect to classical ones than a scenario without an additional planet \citep{Gomes2015}. Moreover, \cite{Gomes2017} (see also \cite{Bailey2016} and \cite{Lai2016}) provide an explanation for the current inclination of the planetary system invariant plane with respect to  the Sun's equator of $\sim6^\circ$. They arrived at parameters for the distant planet compatible with those found by \cite{Brown2016} although with possible larger eccentricities corresponding to a given semimajor axis.

As for the proposed mechanisms for its origin, Planet 9 could have been originally dispersed from the giant planets region. In order to avoid that repeated close encounters with a giant planet could end in the ejection of the scattered planet, the perihelion distance of its orbit would have to be lifted by some mechanism. These might include its gravitational interaction with the Sun's birth cluster (\cite{Brasser2006}, \cite{Brasser2012}), dynamical friction due to a planetesimals disk located beyond $100$ au \citep{Eriksson2018} or an either massive and short-lived or low-mass and long-lived gas disk \citep{Bromley2016}. Other proposed scenarios suggest the capture of a planet from another star system or the capture of a free-floating planet in the solar birth cluster (\citet{Li2016}, \citet{Mustill2016}; see, however, \citet{Parker2017}).

\citet{Brown2016} present tables with probabilities for a Planet 9 with a given semimajor axis and eccentricity to induce a favorable orbital confinement of large semimajor axis TNOs. One of their criteria to choose the best Planet 9's orbit is to rule out those that produce too many high perihelion and low semimajor axis TNOs. Here we analyze the evidence of a distant planet, without neglecting the case of planets with low perihelion that were naturally neglected in \citet{Brown2016}, since they would assumedly produce an excess of high perihelion TNOs for relatively small semimajor axes. We particularly analyze these low perihelion Planet 9's influence on the Kuiper Belt and on the relation between scattered and detached objects. In Section \ref{methods} we describe the methods used to test the influence of Planet 9 on large semimajor axis TNOs. We present our results in Section \ref{results}. In Section \ref{lowperihelion} we emphasize the case of low perihelion planets. In Section \ref{conclusions} we present our conclusions.

\section{Methods}\label{methods}

The origin of the detached objects (defined as minor bodies with $a>50$ au and $q>40$ au) with $a\lesssim200$ au can be explained as a result of the increase of the perihelion distance of scattered disk objects (bodies with $a>50$ au and 30 au $<q<$ 40 au) via resonant capture with Neptune associated with the Kozai mechanism. These objects may still be experiencing those resonances or they may have been released from the resonant coupling MMR-Kozai during the migration of Neptune after the instability phase of the Solar System (\cite{Gomes2005}, \cite{Morbidelli2005}, \cite{Tsiganis2005}), when their orbital eccentricities were low enough (\cite{Gomes2005b}, \cite{Gomes2011}, \cite{Brasil2014}).

\begin{figure}[ht!]
\epsscale{1.1}
\plotone{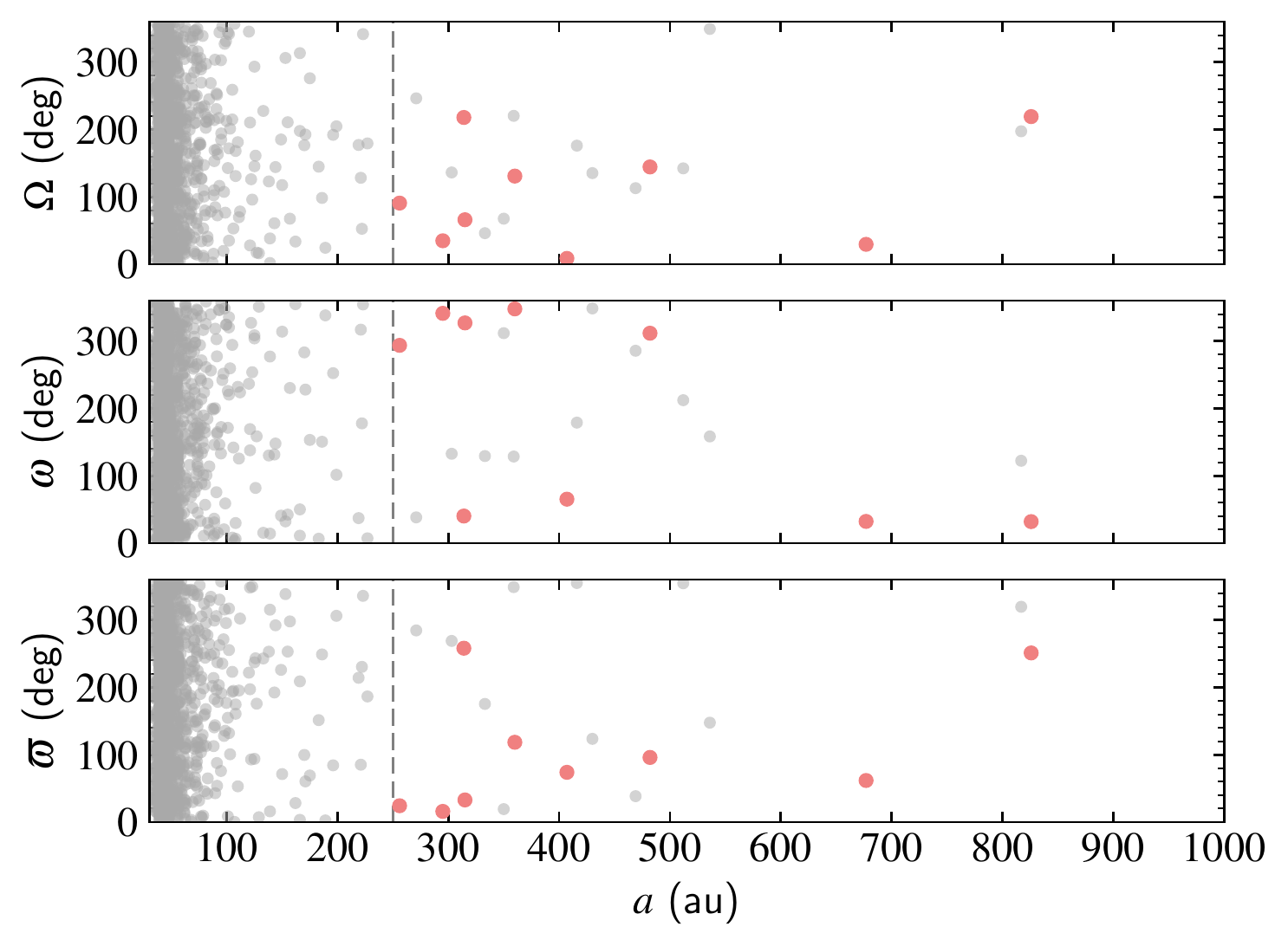}
\caption{Longitude of the ascending node, argument of perihelion and longitude of perihelion as a function of the semimajor axis for all trans-Neptunian objects observed in more than one opposition and up to $a=1000$ au, those with $a\geq250$ au and $q\geq40$ au are highlighted in red.\label{Distant_TNOs}}
\end{figure}

Our analysis is motivated by  the trans-Neptunian objects with $q\geq40$ au and $a\geq250$ au, whose existence can not be explained by the current orbital configuration of the giant planets. Currently, there are nine\footnote{As of May 2018.} objects within this category observed in more than one opposition in the Minor Planet Center (MPC) database\footnote{\url{https://minorplanetcenter.net/iau/lists/Centaurs.html}}. As shown in Figure \ref{Distant_TNOs} they have $\Omega<220^\circ$, a confinement of $131.7^\circ$ in $\omega$ (between $\omega=293.6^\circ$ and $\omega=65.3^\circ$), and two likely confinements in $\varpi$, where the main confinement, of $102.6^\circ$, is composed of 7 objects and located around $\varpi=60.5^\circ$, and the secondary one, represented by the two remaining objects, around $\varpi=254.6^\circ$. These objects have $q<81$ au and $i<26^\circ$. We decided to neglect objects with smaller semimajor axis and perihelion distance than used in other works (e.g. \cite{Batygin2016}) concerning Planet 9 since the largest $a$ and $q$ are, the likeliest are that these objects have not experienced an important close encounter with Neptune in a fairly recent past. Thus these objects are prone to keep the angular alignment induced by Planet 9. We note that considering distant TNOs with these higher lower boundaries in semimajor axis and perihelion distance, and also the existence of a secondary clustering, the angular confinement is more remarkable for $\varpi$ and $\omega$ than for $\Omega$. Surveys carried out near the ecliptic plane would favor the detection of objects with $\omega=0^\circ$ or $\omega=180^\circ$. The fact that there is no confinement around $\omega=180^\circ$ and that several of these objects were discovered in off-ecliptic surveys argue that the confinement around $\omega=0^\circ$ is not an effect due to observational bias (\cite{Trujillo2014}). It should be noted, however, that even though the clustering in $\omega$ observed near $0^{\circ}$ is not an artifact of observational bias, a real clustering of $\omega$ near $0^{\circ}$ will be observationally intensified due to the bias above mentioned.

We performed numerical integrations for the age of the Solar System considering different combinations of perihelion distance, inclination and argument of perihelion for two semimajor axes of Planet 9, $a_9=700$ au and $a_9=1500$ au, where for the models with the small semimajor axis we consider a mass of $3\times10^{-5}M_{\odot}$, and a mass of $5\times10^{-5}M_{\odot}$ for the other case. The tested grids are shown in the first column of Table \ref{table1}. The nomenclature for each model means $a_9$-$i_9$-$\omega_9$-$q_9$-$m_9$, where the semimajor axis and perihelion distance are in au, the inclination and argument of perihelion in degree, and the mass in $10^{-5}$ M$_{\odot}$ ($\sim 3.3$ M$_{\oplus}$).

The initial conditions are given in relation to the plane perpendicular to the resultant angular momentum vector of the four known giant planets (hereinafter referred to as iv4) in their current configuration, which defines the initial orbital elements for the known giant planets. Besides these, the system is made up also of an external perturber and 10,000 test particles\footnote{For one of the cases (planet with $a_9=700$ au and orbital inclination at $10^{\circ}$) we considered $20,000$ particles since there were too few particles in the end of the integration to perform a reliable statistics.} with orbital elements similar to those that would be expected some time after the instability phase of the giant planets under the Nice Model (\cite{Gomes2005}, \cite{Morbidelli2005}, \cite{Tsiganis2005}): 200 au $<a<$ 1000 au, 30 au $<q<$ 40 au, $0^\circ<i<10^\circ$ and the remaining orbital elements uniformly distributed. The lower limit for the particles' semimajor axes stands for the fact that we are only interested in large semimajor axis TNOs to better characterize their confinement. Due to the torque exerted by the ninth planet in an inclined orbit on known planets the initial reference plane will precess with respect to the invariable plane of the system, so that our outputs are converted relative to the instantaneous iv4 plane.

The integrations were done with the MERCURY package hybrid integrator \citep{Chambers1999}. To profit from a multi-core supercomputer, for each model we first ran an integration with just the planets, saving their coordinates at each $10$ years for later use. After that we performed one hundred numerical integrations, each in a different core, reading the coordinates from a file produced by the run with just the planets. These integrations with particles were done normally with the same timestep of 0.5 year, but at every 10 years we corrected back the planets coordinates to those produced by the original integration with no particles. This is a quite easy and reliable procedure. The deviation of the planets after 10 years (20 timesteps) is always quite small since the particles are massless and the difference between the integration with particles and the original one with just the planets comes solely by the fact that the (hybrid) integrator can enter into the classic integration mode (bulirsch-stoer) at close encounters of particles with planets. We thus modified the code for the hybrid integrator to implement this procedure. For each time considered in the analyses, and for each particle, we associate that time and ten other neighboring ones, five subsequent ones and five previous ones, entailing a total $\Delta t= 10^6$ yr, in order to improve the statistics. In $10^6$ years the circulation of the angles $\Omega$, $\omega$ and $\varpi$  for the distant TNOs and Planet 9 is quite small so that this procedure does not spoil the investigation of confinements. 

For the simulated particles belonging to a range in $a$, $q$ and $i$ similar to that of the observed objects, our analysis consists of determining the time evolution of i) the highest and second-highest peaks in $\Omega$, $\omega$ and $\varpi$ as determined by histograms of the distribution of those angles, ii) the angles corresponding to the middle of the bins that yielded the highest and second-highest peaks, and iii) the confinements in $\Omega$, $\omega$ and $\varpi$. These are described in detail below.

\begin{figure}[ht!]
\epsscale{1.1}
\plotone{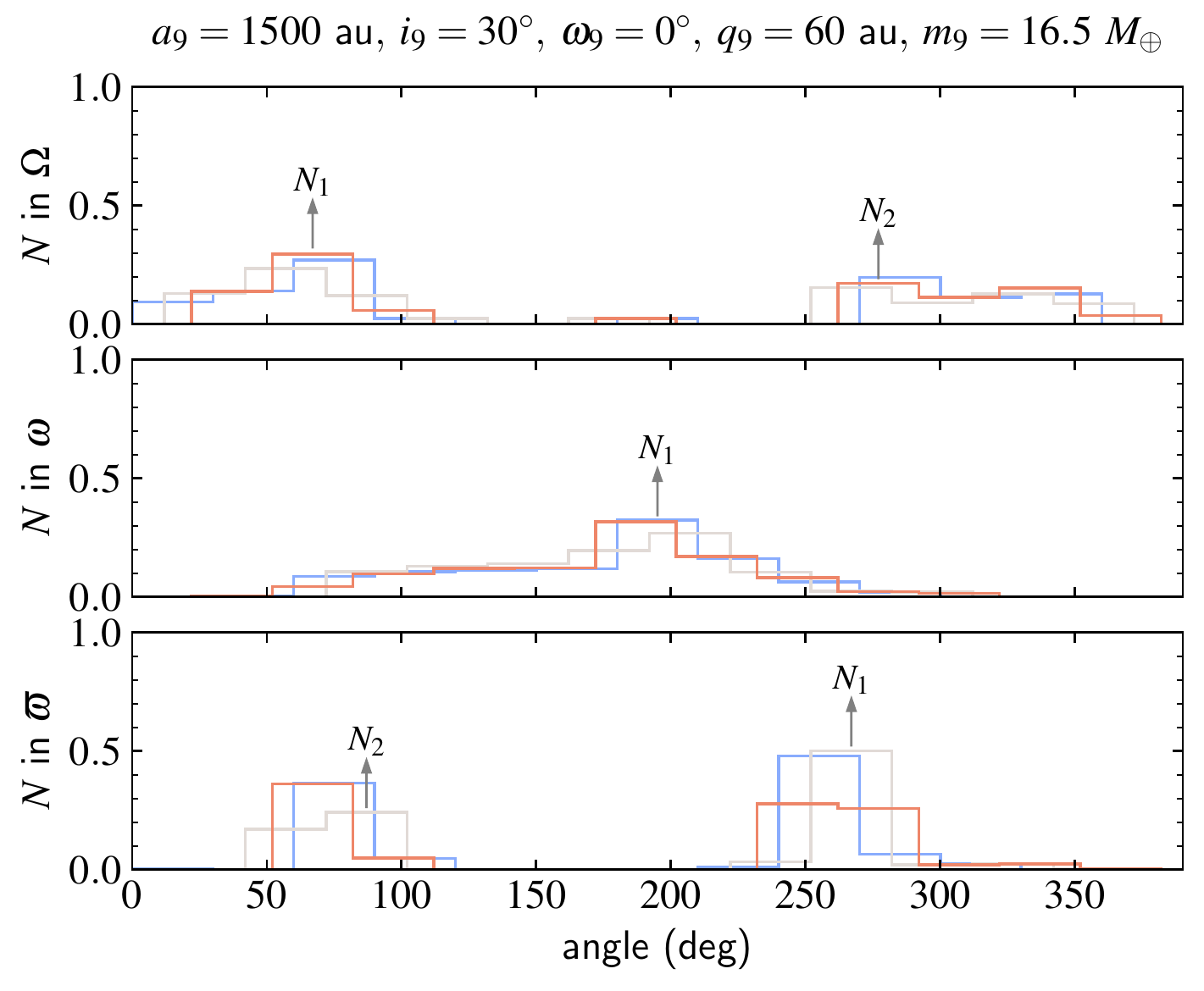}
\caption{Frequencies in $\Omega$, $\omega$ and $\varpi$ for the model a1500-30-0-60-5 at a time $t = 2.18$ Gyr showing $3$ different beginnings of the first bin at $0^\circ$, $12^\circ$ and $22^\circ$ in blue, gray and red, respectively; out of a total of 30 considered in each time to calculate the highest and secondary frequencies in these orbital elements. For $\Omega$ and $\varpi$ we can distinguish a principal and a secondary confinement, which are distant from one another of $\sim180^\circ$. There is no secondary frequency in $\omega$.\label{amostra_hist}}
\end{figure}

The highest and second-highest peaks are determined for each time by performing histograms of the number of particles that have their $\Omega$, $\omega$ and $\varpi$ inside the histograms' bins. These particles are a subset of the simulated particles whose semimajor axes, perihelion distances and inclinations belong to the interval 250 au $\leq a\leq$ 1000 au, 38 au $\leq q\leq$ 90 au and $0^\circ\leq i\leq30^\circ$. The absolute number of particles in each bin is transformed into frequencies, $N$, given by the number of particles associated to that bin divided by the total number of particles in all bins. The bins are $30^{\circ}$ wide so there is a total of $12$ bins. Thus a flat distribution yields a frequency of $1/12$ for each bin. The idea of choosing a bin $30^{\circ}$ wide is based on visual inspection of the clustering of some angles during the evolution of the simulated TNOs. We saw that this clustering can be quite narrow and $30^{\circ}$ was chosen so as to capture this peculiar feature. For each time we make $30$ histograms, which differ from each other in the initial angle, varying from $0^{\circ}$ to $29^{\circ}$, so that we can more accurately determine the bin where there is the greatest clustering of angles, which is the one with highest $N$ in all $30$ histograms for the same time. Figure \ref{amostra_hist} depicts the construction of the histograms for $\Omega$, $\omega$ and $\varpi$ for three different initial positions of the bins. In both  $\Omega$ and $\varpi$ we see the appearance of two peaks in the angular frequencies. Unless the distribution is totally flat, which is very unlikely, there is always a bin with a highest frequency. The secondary frequency is defined whenever there is a secondary peak in the histogram. It is not the second highest frequency in the histogram which would often be associated just to a bin next to the highest frequency. Note that we are thus obtaining the values corresponding to two confinements with a minimum angular distance from each other of $60^\circ$. In this way, we get the highest ($N_1$) frequency and the respective secondary one ($N_2$) for each time and their respective associated angles. 

The idea behind this analysis is that a bin with a high frequency corresponds to a relevant confinement in the angular interval defined by the bin. The measurement of the secondary frequency is motivated not only from the observational data shown in Fig. \ref{Distant_TNOs} where the distribution of $\varpi$ suggests the existence of two confinements, but also on visual inspection of the evolution of the angular confinements produced by the simulations.

We also perform a confinement test similar to that in \citet{Brown2016}. For each time, we also consider the particles that belong to the interval 250 au $\leq a\leq$ 1000 au, 38 au $\leq q\leq$ 90 au and $0^\circ\leq i\leq30^\circ$ and we do one hundred iterations, where in each iteration we randomly select $9$ objects from the set of particles (because we are considering $9$ observed distant TNOs); for $\Omega$ and $\omega$ we calculate the confinement for the 9 objects and for $\varpi$ we determine the smallest possible confinement for 7 of those 9 objects (since, as mentioned above, we are assuming that there are two confinements in $\varpi$). Thereby, we obtain for each time the average and standard deviation of the confinements.

\section{Results}\label{results}

\begin{figure}[ht!]
\epsscale{1.1}
\plotone{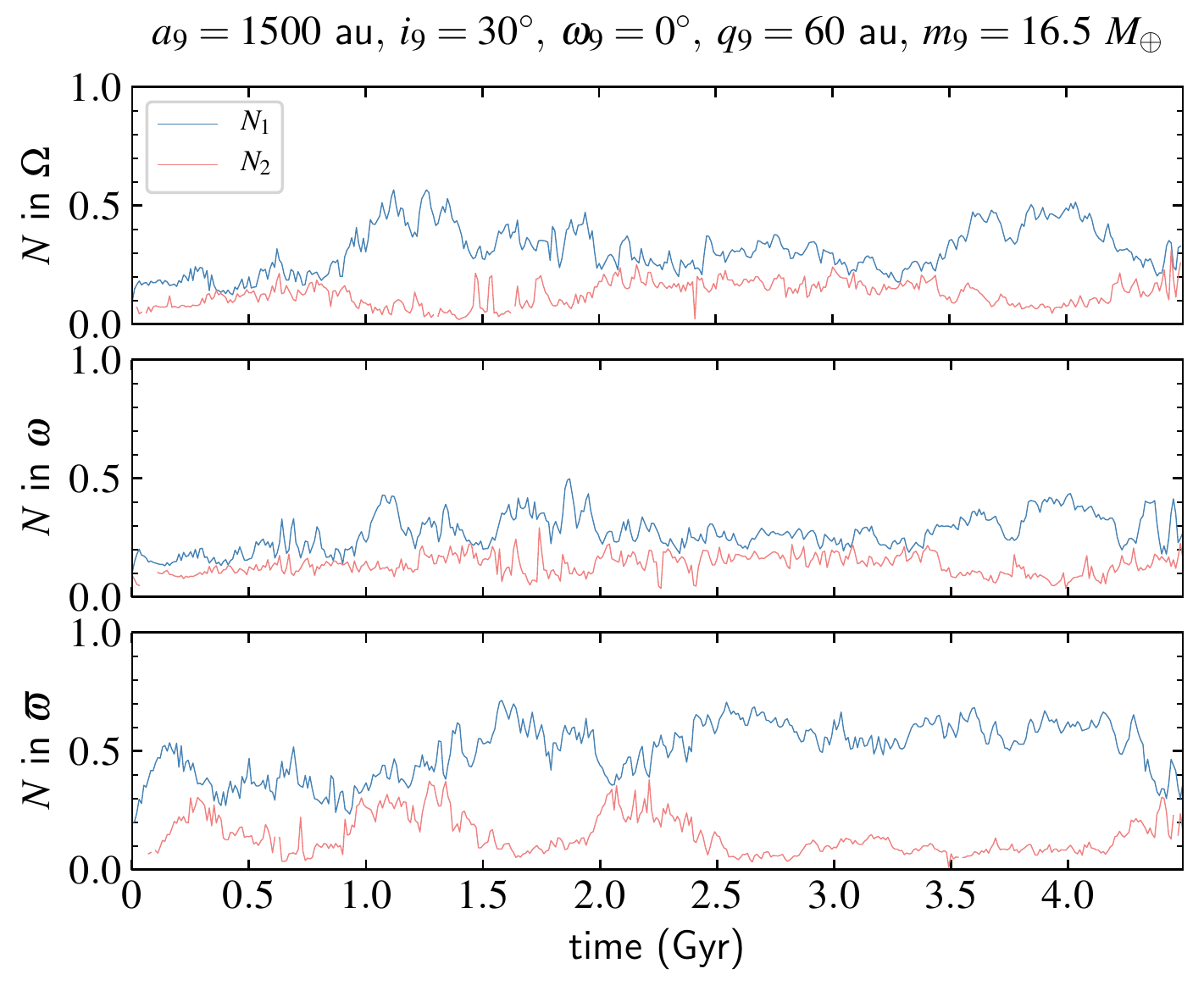}
\caption{First and secondary peaks in the angular frequencies as a function of time shown with blue and red, respectively for the model $a$1500-30-0-60-5.\label{a1500-30-0-60-5_freqrel_vs_t}}
\end{figure}

\begin{figure}[ht!]
\epsscale{1.1}
\plotone{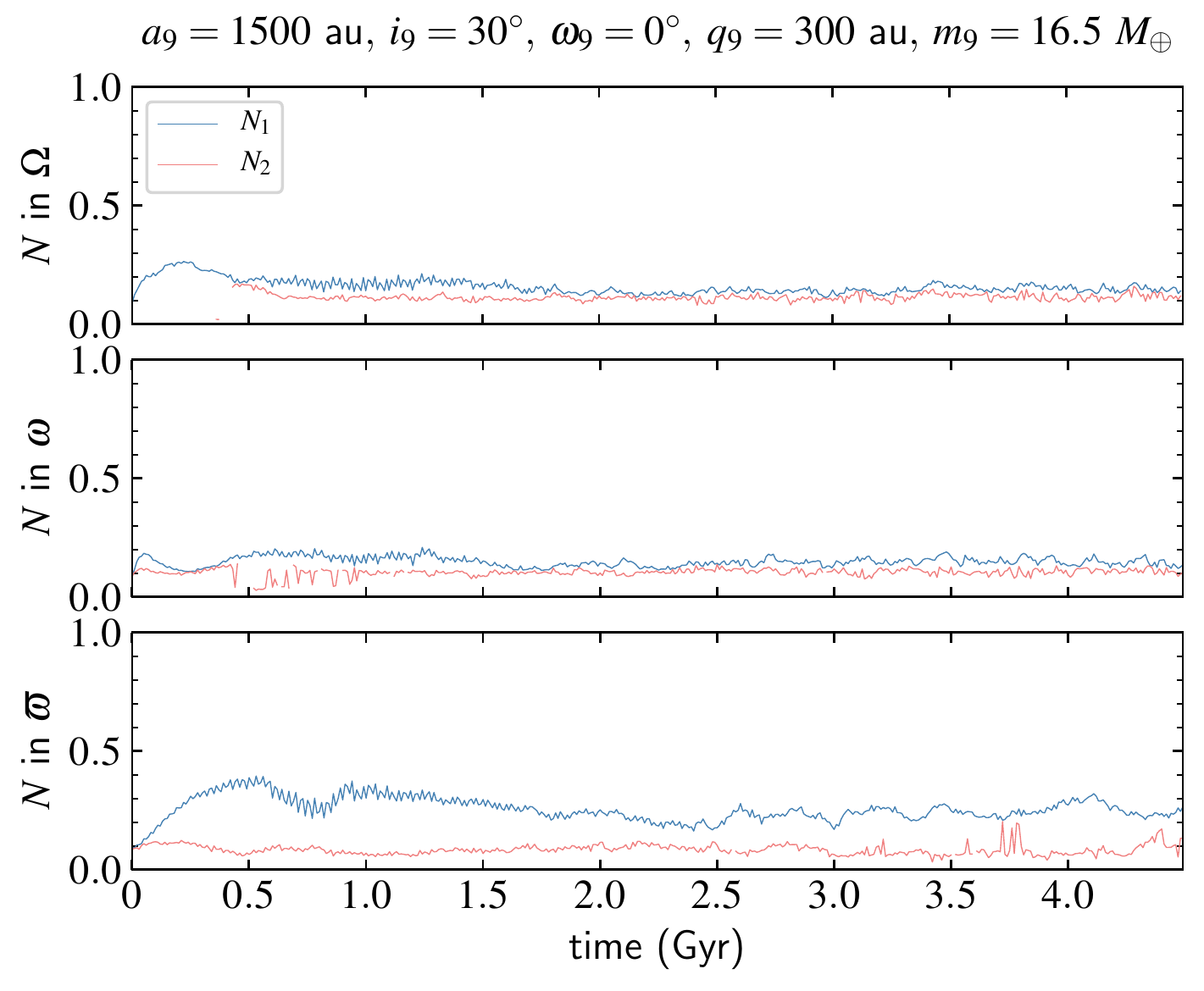}
\caption{Same as Fig. \ref{a1500-30-0-60-5_freqrel_vs_t} for the model $a$1500-30-0-300-5. The values of the highest and secondary frequencies are quite low, mainly in $\Omega$ and $\omega$ where both primary and secondary frequencies have similar low values implying that there is no well defined confinement.\label{a1500-30-0-300-5_freqrel_vs_t}}
\end{figure}

The time evolutions of the highest and secondary frequencies in the angles $\Omega$, $\omega$ and $\varpi$ obtained from histograms show better defined angular confinements during the first billion years in almost all our models. Figures \ref{a1500-30-0-60-5_freqrel_vs_t} and \ref{a1500-30-0-300-5_freqrel_vs_t} present the results for two extreme cases of perihelion distances, $q_9=60$ au and $q_9=300$ au, both for semimajor axis of Planet 9 of 1500 au. These figures suggest that higher frequencies, therefore, narrower confinements, are usually associated to smaller perihelion distances of Planet 9. This trend is also true for models with $a_9=700$ au.

\begin{figure}[ht!]
\epsscale{1.1}
\plotone{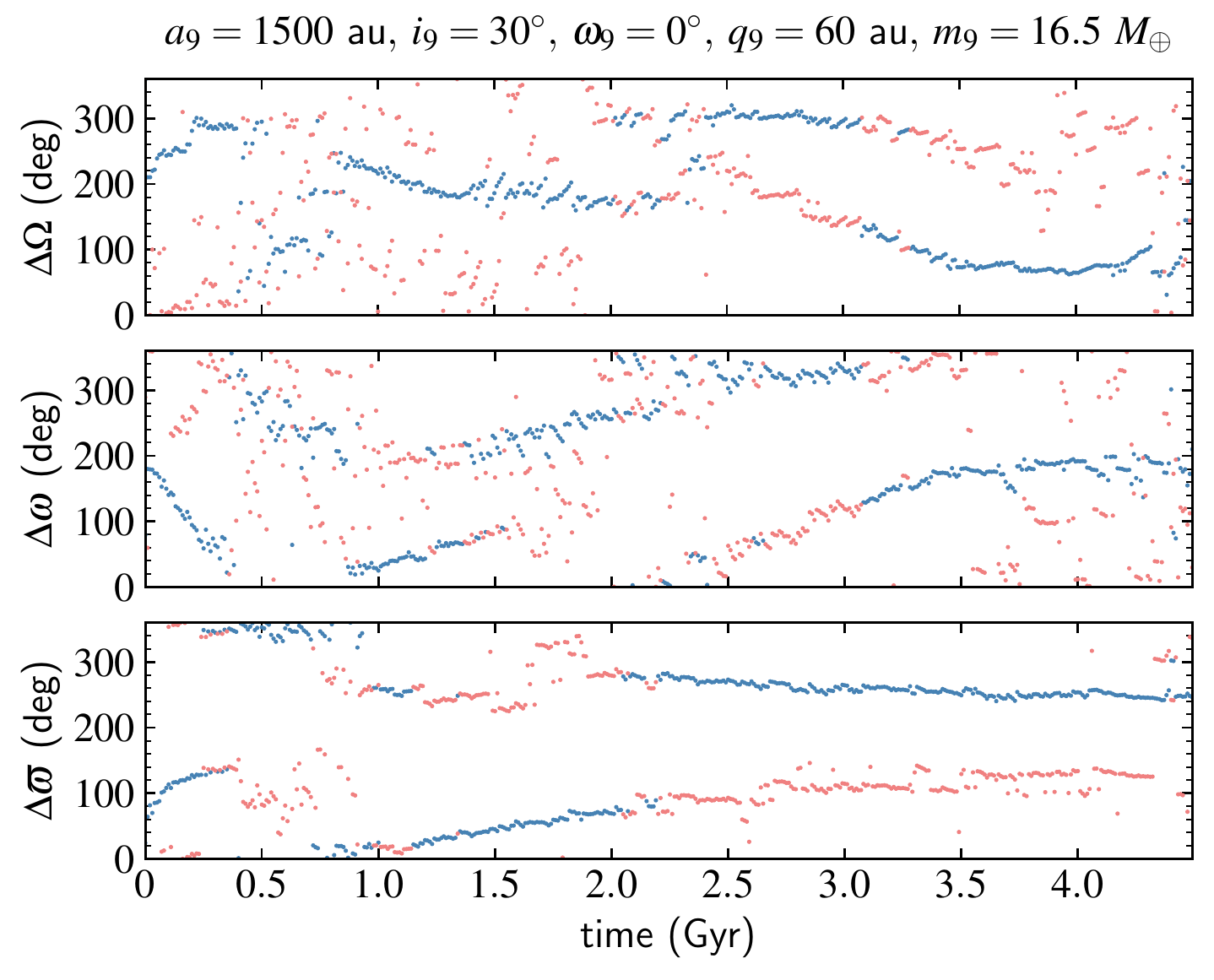}
\caption{Evolution of the angular elements of the particles' orbits in relation to those of the Planet 9 corresponding to the highest and secondary frequencies shown with blue and red colors respectively, for the model $a$1500-30-0-60-5. For $\varpi$, the main and secondary confinements exchange positions several times until $\sim 2$ Gyr, when the main confinement is kept roughly unchanged. \label{a1500-30-0-60-5_Dangles_vs_t}}
\end{figure}

\begin{figure}[ht!]
\epsscale{1.1}
\plotone{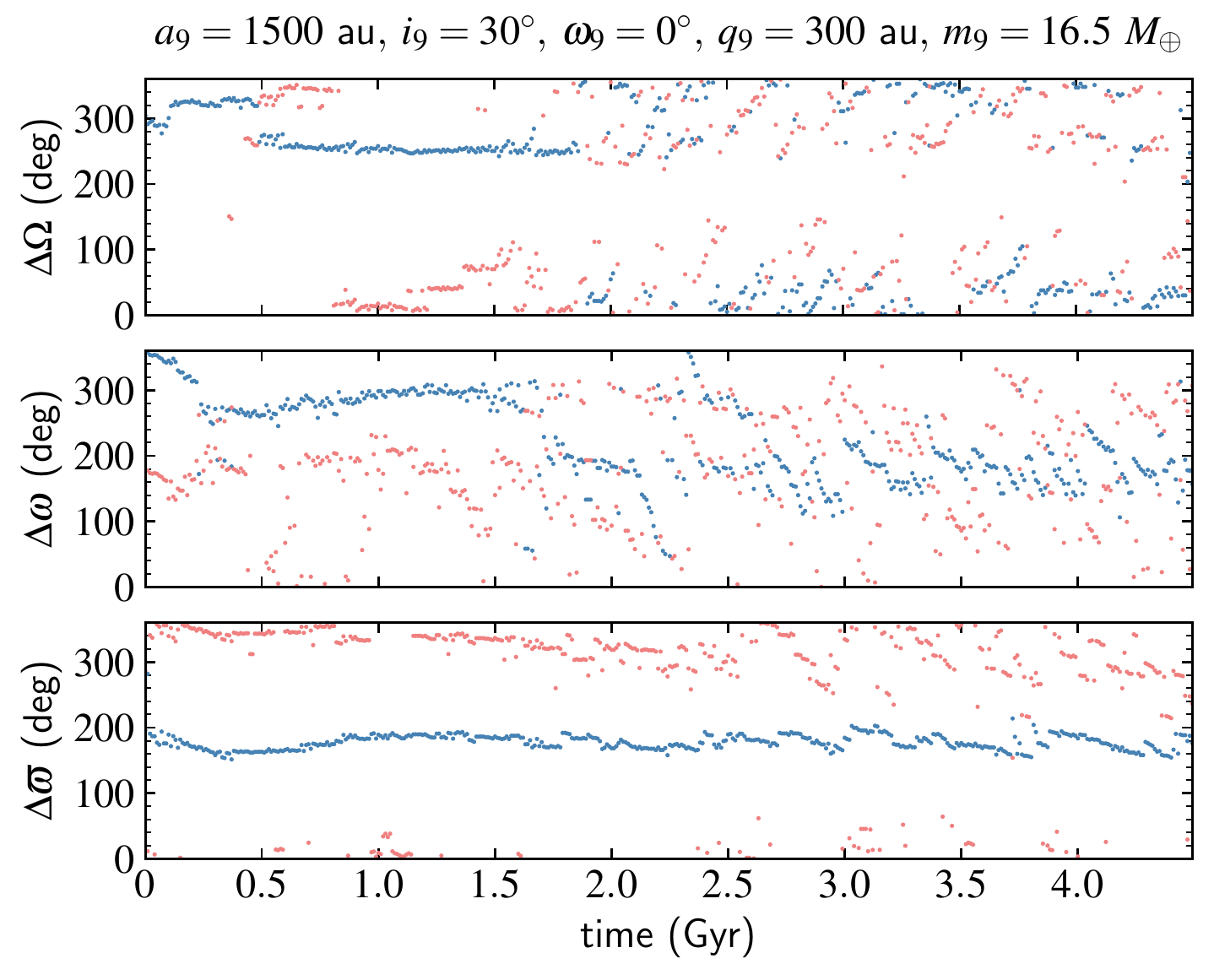}
\caption{Same as Fig. \ref{a1500-30-0-60-5_Dangles_vs_t} for the model $a$1500-30-0-300-5. The particles do not maintain an approximately constant relative distance for the last billion years in neither $\Delta\Omega$ nor $\Delta\omega$, as expected for great perihelion distances of the planet 9.\label{a1500-30-0-300-5_Dangles_vs_t}}
\end{figure}

The angles corresponding to the highest and second-highest peaks in the frequencies are calculated in relation to that of Planet 9 ($\Delta\Omega$, $\Delta\omega$ and $\Delta\varpi$). In most of our models Planet 9 leads to a shepherding in $\varpi$ of the set of more tightly confined particles (those within the angular range corresponding to the highest frequency) and in several of our models this is also true for the particles belonging to the second confinement. Figure \ref{a1500-30-0-60-5_Dangles_vs_t} shows an example of this behavior, where the angular distance between both confinements in $\varpi$ is not constant and at $4.5$ Gyr it is $\sim100^\circ$. We also notice that the highest and secondary frequencies switch positions several times. When we compare this with Fig. \ref{a1500-30-0-60-5_freqrel_vs_t} we see that this switching happens when the highest and secondary frequencies have similar values. For $\Omega$ and $\omega$ the angular distance between the particles' center of angular clustering and the corresponding angle of Planet 9 is not so well defined in time, but in some of our models the relative distances $\Delta\Omega$ and $\Delta\omega$ corresponding to the highest frequency also remain constant for about the last Gyr. Models with large perihelion distances show a large dispersion in the relative angles (Figure \ref{a1500-30-0-300-5_Dangles_vs_t}).  

Table \ref{table1} shows the average and standard deviation taken for the last $0.5$ Gyr of simulation time, of the angular distance between the center of angular clustering and the corresponding planetary angle. We see that for $\varpi$ this distance varies inside the range $172.4^{\circ}$ - $248.9^{\circ}$, thus giving a dispersion of $76.5^{\circ}$. The average of this angular distance is $206.6^{\circ}$, which suggests the likely position of Planet 9' longitude of perihelion at $\sim 267^\circ$. For $\Omega$ and $\omega$ this angular distance is not well defined, but varies roughly uniformly around all $360^{\circ}$ range. This is true even if we just consider the cases of smaller standard deviations where the clustering position is better defined.

\begin{table}[h!]
\centering
\caption{Mean and standard deviation (std) of the angular distance between the particles' primary center of clustering and the same angle for Planet 9, computed for the last 500 Myr.} \label{table1}
\setlength\tabcolsep{3.7pt} 
\begin{tabular}{l r r r r r r}
\tablewidth{0pt}
\hline
\hline
\multirow{2}{*}{Model} & \multicolumn{2}{c}{$\Delta\Omega$ (deg)} & \multicolumn{2}{c}{$\Delta\omega$ (deg)} & \multicolumn{2}{c}{$\Delta\varpi$ (deg)}\\ \cmidrule(lr){2-3} \cmidrule(lr){4-5} \cmidrule(lr){6-7}
& Mean & Std & Mean & Std & Mean & Std\\
\hline
\decimals
$a$1500-10-0-100-5 & 107.0 &   7.0 &  73.9 &  21.7 & 180.2 &  18.7\\
   $a$1500-30-0-60-5 &  78.4 &  18.5 & 181.8 &  17.7 & 248.9 &   4.4\\
  $a$1500-30-0-100-5 & 127.4 &  15.8 & 252.5 &  13.2 & 232.2 &   2.4\\
  $a$1500-30-0-200-5 & 307.0 &  93.5 & 265.6 &  75.8 & 174.6 &  19.3\\
  $a$1500-30-0-300-5 & 351.2 &  55.1 & 185.1 &  28.2 & 172.4 &  10.8\\
  $a$1500-30-90-60-5 & 211.4 &  75.7 & 216.2 &  82.5 & 184.8 &  85.4\\
  $a$1500-60-0-100-5 & 265.5 &  12.8 & 305.6 &  17.9 & 201.8 &   5.4\\
   $a$700-10-0-100-3 & 193.5 &  68.4 &  58.7 &  98.8 & 233.8 &   9.2\\
    $a$700-30-0-60-3 & 112.9 &  77.5 & 182.5 &  71.0 & 182.1 &  62.0\\
   $a$700-30-0-100-3 & 125.9 &  71.0 & 255.6 &  66.5 & 191.4 &  13.8\\
   $a$700-30-0-200-3 & 316.0 &   7.1 & 266.7 &   7.0 & 223.1 &   4.1\\
   $a$700-30-0-300-3 & 148.5 & 100.1 & 227.2 &  81.2 & 239.6 &  27.1\\
   $a$700-30-90-60-3 & 244.3 &  68.9 & 333.5 &  84.3 & 229.1 &  19.6\\
   $a$700-60-0-100-3 & 100.9 &  82.2 & 142.1 &  75.3 & 198.1 &  21.5\\
\hline
\end{tabular}
\end{table}

\begin{figure}[ht!]
\epsscale{1.1}
\plotone{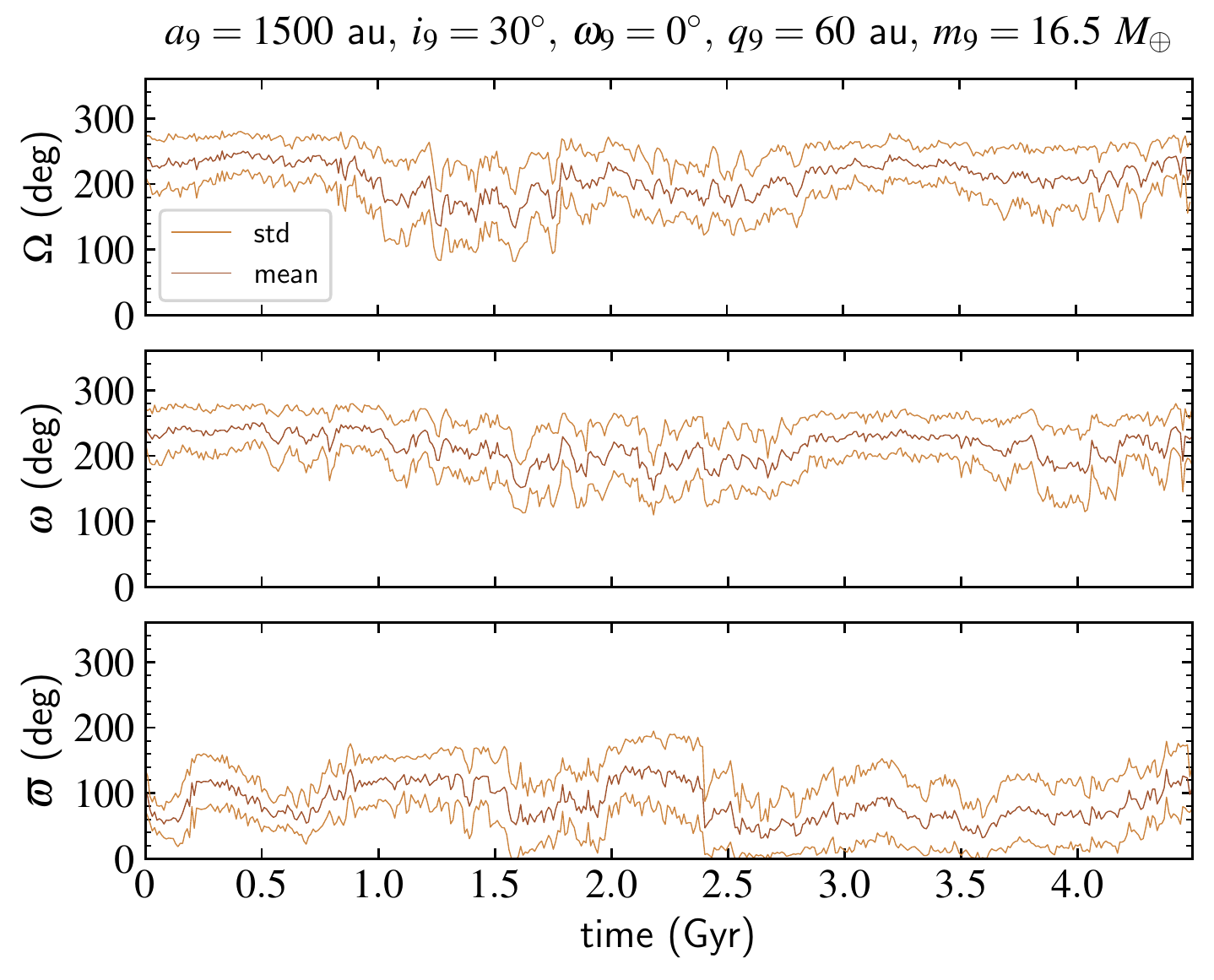}
\caption{Confinement mean and its standard deviation in $\Omega$, $\omega$ and $\varpi$ as a function of time shown with dark and light brown, respectively for the model $a$1500-30-0-60-5.\label{a1500-30-0-60-5_conf_vs_t}}
\end{figure}

\begin{figure}[ht!]
\epsscale{1.1}
\plotone{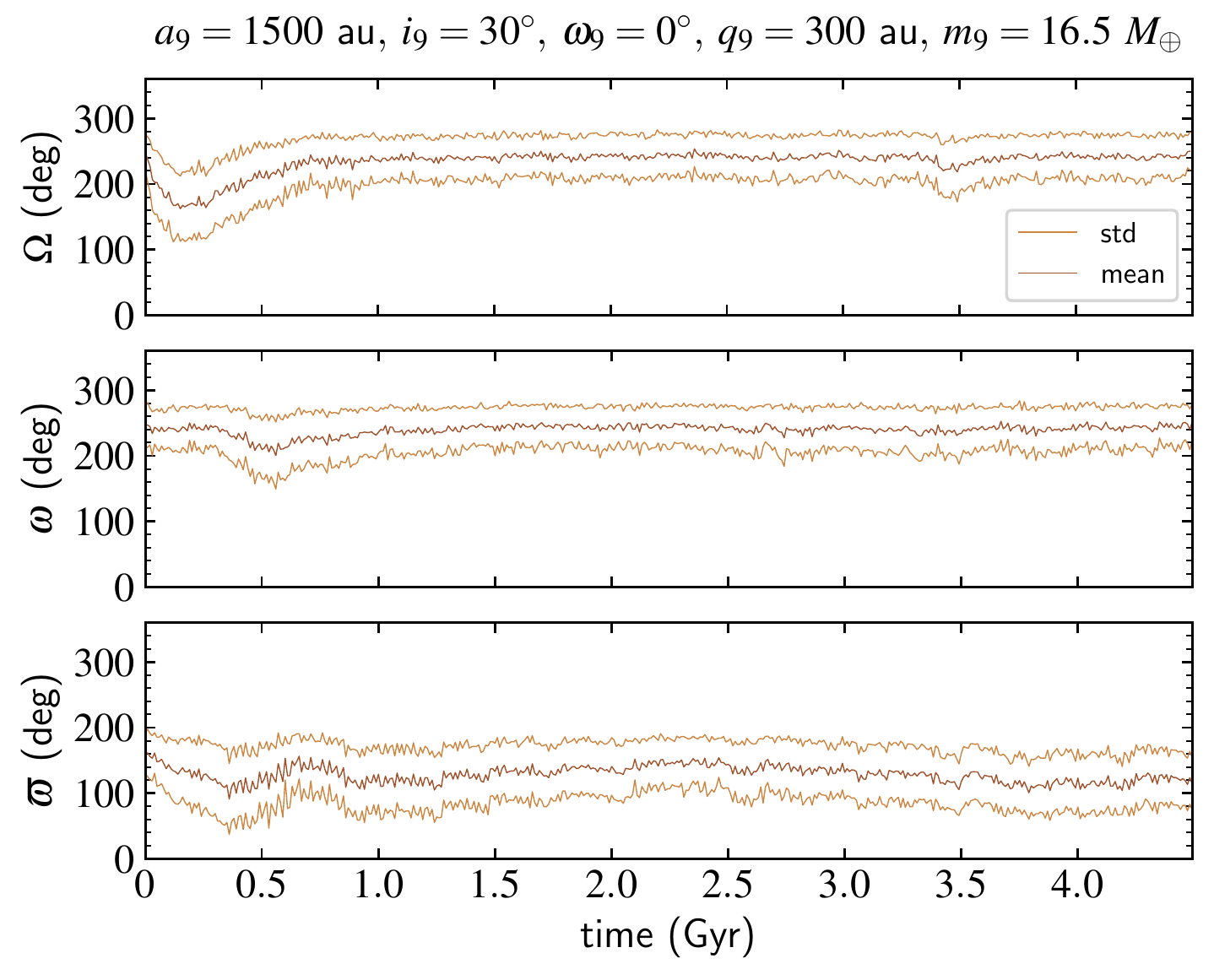}
\caption{Same as Fig. \ref{a1500-30-0-60-5_conf_vs_t} for the model $a$1500-30-0-300-5. From about 1 Gyr results are very similar to those that would be expected in the case of a random distribution.\label{a1500-30-0-300-5_conf_vs_t}}
\end{figure}

Our analysis of confinement as a function of time confirms our previous results. In all models a better confinement for the particles is reproduced for $\varpi$ than for $\Omega$ or $\omega$ and planets with smaller perihelion distances are associated with better confinements (compare Figs. \ref{a1500-30-0-60-5_conf_vs_t} and \ref{a1500-30-0-300-5_conf_vs_t}). For models with $q_9\sim300$ au, the confinements, mainly in $\Omega$ and $\omega$, are near to what would be expected in the case of a randomly produced uniform distribution in these angular elements when the procedure indicated in the Section \ref{methods} is applied.

\begin{figure}[ht!]
\epsscale{1.1}
\plotone{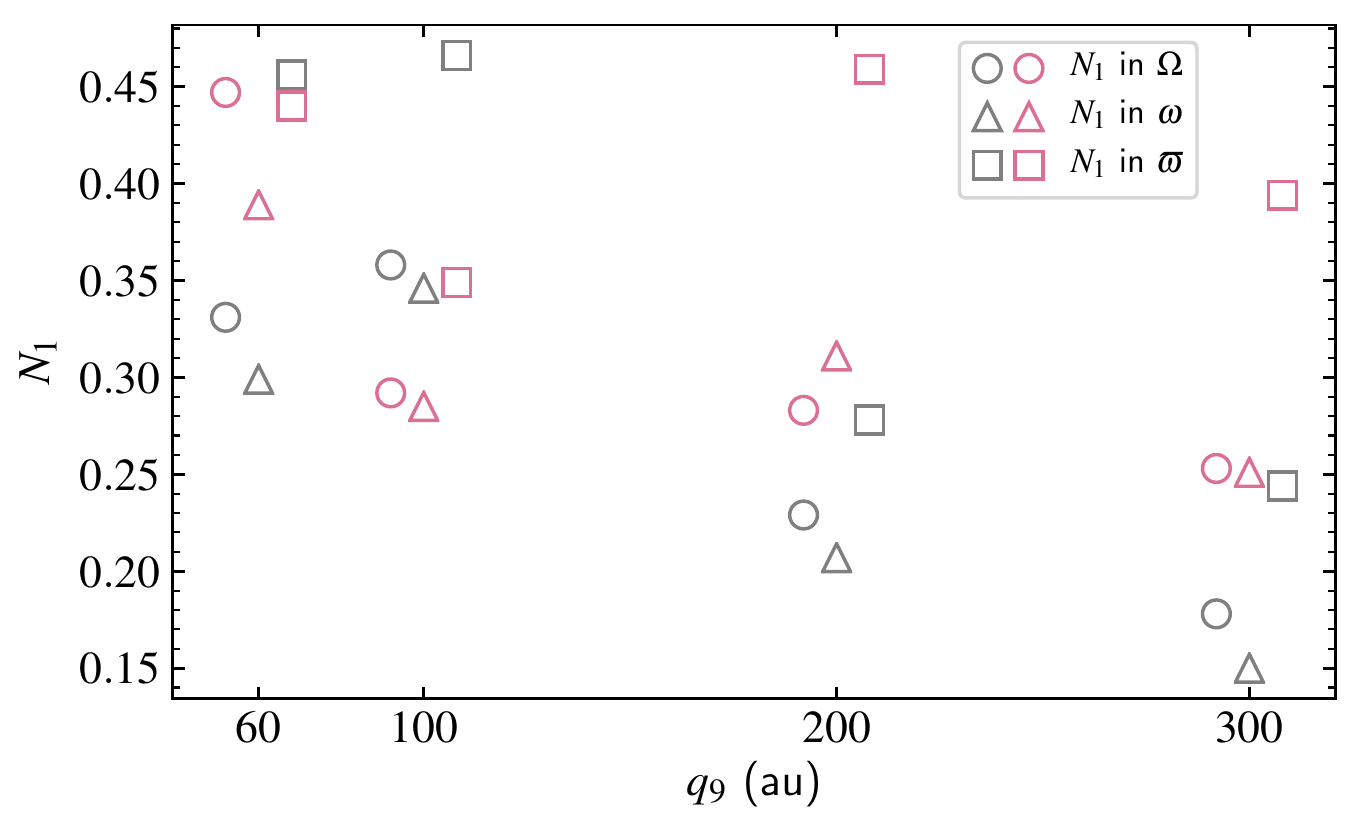}
\caption{Mean values of $N_{1}$ in $\Omega$, $\omega$ and $\varpi$ (represented by circles, triangles and squares respectively) calculated for the complete integration time for models with $i_9=30^\circ$ and $\omega_9=0^\circ$. The gray color represents models with $a_9=1500$ au and the violet one those with $a_9=700$ au.\label{tabela2_fig}}
\end{figure}

In Table \ref{table2} we depict the maximum and mean values of the collection of highest frequencies (main confinements) in $\Omega$, $\omega$ and $\varpi$ of the particles considering the full integration time (see also Fig. \ref{tabela2_fig}) and the last 500 Myr, for all our models. The reliability of the numbers presented in Table \ref{table2} and plotted in several figures depend on the number of points used to build the statistics for each time. Since particles are continually ejected by the planets during the simulation, the worst cases are always near the end of the integrations. Moreover, planets with lower perihelion distances eject more particles through the whole evolution than those with higher perihelion distances. In most of the cases, and for any time, we have anyway always more than 50 points to build the statistics. For one specific case, as commented in Section \ref{methods}, we doubled the number of points. For that case, there were several times with less than 20 points and we assessed that there were not enough points to produce reliable numbers for the end of the simulation (the last $0.5$ Gyr). As to the other cases that were not doubled, the worst one is for Planet 9 with $a_9=700$ au, $i_9=30^\circ$, $\omega_9=0^\circ$ and $q_9=60$ au, for which we have 24 times with less than 50 points, but just two times with less than 20 (17 and 16 points). Also since part of results shown in Table \ref{table2} are given in averages for at least the last 0.5 Gyr, even a small randomization due to small numbers can be averaged out. For planetas with the lowest perihelion distances (60 and 100 au), however, one must consider with caution the results described by maxima.

In general,  lower perihelion distances of Planet 9 produce better defined confinements in the distant TNOs. However, a constraint that an outer planet with a low perihelion distance should satisfy is the preservation of the Classical Kuiper Belt. Another important constraint refers to the right ratio of observed detached to scattering objects, since a distant planet tends to overpopulate the region of the detached objects. These will be the subjects of the next section.

\begin{table*}[h!]
\centering
\caption{Maximum and mean values of the angular frequencies corresponding to the narrowest confinements calculated considering both the 4.5 Gyr and the last 500 Myr of integration time.} \label{table2}
\setlength\tabcolsep{7.3pt} 
\begin{tabular}{l l l l l l l l l l l l l}
\tablewidth{0pt}
\hline
\hline
\multirow{3}{*}{Model} & \multicolumn{4}{c}{$N_1$ in $\Omega$} & \multicolumn{4}{c}{$N_1$ in $\omega$} & \multicolumn{4}{c}{$N_1$ in $\varpi$} \\ \cmidrule(lr){2-5} \cmidrule(lr){6-9} \cmidrule(lr){10-13}
 & \multicolumn{2}{c}{4.5 Gyr} & \multicolumn{2}{c}{last 500 Myr} & \multicolumn{2}{c}{4.5 Gyr} & \multicolumn{2}{c}{last 500 Myr} & \multicolumn{2}{c}{4.5 Gyr} & \multicolumn{2}{c}{last 500 Myr} \\ 
 & max & mean & max & mean & max & mean & max & mean & max & mean & max & mean \\
\hline
  $a$1500-10-0-100-5 &  0.362 &  0.226 &  0.219 &  0.183 &  0.313 &  0.202 &  0.194 &  0.164 &  0.472 &  0.283 &  0.376 &  0.319\\
   $a$1500-30-0-60-5 &  0.567 &  0.331 &  0.515 &  0.359 &  0.499 &  0.299 &  0.436 &  0.308 &  0.714 &  0.456 &  0.665 &  0.483\\
  $a$1500-30-0-100-5 &  0.621 &  0.358 &  0.304 &  0.236 &  0.595 &  0.346 &  0.337 &  0.266 &  0.749 &  0.466 &  0.651 &  0.545\\
  $a$1500-30-0-200-5 &  0.367 &  0.229 &  0.196 &  0.163 &  0.323 &  0.207 &  0.219 &  0.175 &  0.464 &  0.278 &  0.269 &  0.241\\
  $a$1500-30-0-300-5 &  0.265 &  0.178 &  0.175 &  0.152 &  0.210 &  0.150 &  0.182 &  0.151 &  0.395 &  0.244 &  0.321 &  0.266\\
  $a$1500-30-90-60-5 &  0.397 &  0.244 &  0.354 &  0.259 &  0.405 &  0.251 &  0.344 &  0.262 &  0.493 &  0.293 &  0.318 &  0.248\\
  $a$1500-60-0-100-5 &  0.394 &  0.242 &  0.332 &  0.255 &  0.326 &  0.209 &  0.220 &  0.176 &  0.425 &  0.259 &  0.383 &  0.348\\
   $a$700-10-0-100-3 &  0.848 &  0.473 &  0.848 &  0.530 &  0.681 &  0.386 &  0.494 &  0.336 &  0.768 &  0.486 &  0.651 &  0.456\\
    $a$700-30-0-60-3 &  0.803 &  0.447 &  0.803 &  0.520 &  0.688 &  0.389 &  0.688 &  0.439 &  0.786 &  0.440 &  0.786 &  0.518\\
   $a$700-30-0-100-3 &  0.494 &  0.292 &  0.276 &  0.214 &  0.479 &  0.285 &  0.373 &  0.257 &  0.603 &  0.349 &  0.479 &  0.372\\
   $a$700-30-0-200-3 &  0.470 &  0.283 &  0.470 &  0.324 &  0.525 &  0.311 &  0.458 &  0.313 &  0.744 &  0.459 &  0.707 &  0.577\\
   $a$700-30-0-300-3 &  0.402 &  0.253 &  0.343 &  0.263 &  0.409 &  0.251 &  0.395 &  0.283 &  0.620 &  0.394 &  0.443 &  0.325\\
   $a$700-30-90-60-3 &  0.582 &  0.336 &  0.582 &  0.394 &  0.459 &  0.275 &  0.415 &  0.308 &  0.603 &  0.349 &  0.587 &  0.410\\
   $a$700-60-0-100-3 &  0.832 &  0.465 &  0.553 &  0.372 &  0.766 &  0.429 &  0.653 &  0.422 &  0.882 &  0.534 &  0.632 &  0.458\\
\hline
\end{tabular}
\end{table*}

\section{Other tests for  Planet 9 with low perihelion distance}\label{lowperihelion}

In this section we show the results of the influence of the ninth planet with low perihelion distance on the Classical Kuiper Belt Objects (CKBOs) and on the relation between the Scattered disk and Detached populations within the range 100 au $<a<$ 200 au. 

\subsection{The influence of Planet 9 on the relation between Scattered and Detached objects}

\citet{Brown2016} searched for the best Planet 9 orbits that entail confinements in TNOs angular orbital elements that compares to the confinement encountered in the angular orbital elements of observed TNOs. One of their criteria is to discard those orbital parameters of Planet 9 that lead to the generation of too many particles with perihelion distance $q>42$ au in the range of $100$ au $<a<$ $200$ au, due to the non-existence of such observed objects; namely, they apparently assign a null probability to planets with low $a_9$ and high $e_9$ (low values of $q_9$) since these are prone to produce high perihelion distance objects in that region.

Currently\footnote{As of May 2018.} there are 3 objects (2013 UT$_{15}$, 2014 SS$_{349}$, and 2015 KE$_{172}$, according to the MPC database) observed in more than one opposition with $q>42$ au and $34$ with $q>30$ au, in that range of semimajor axis. We introduce an OBIP (Observational Bias Introducing Procedure) \citep{Gomes2015} to our simulated particles in each model that belong to the range 100 au $<a<$ 200 au and $q>30$ au from 4.4 Gyr and determine the number of objects with $q>42$ au among a sample of 34 objects that can be observed.

OBIP is a procedure that consists of assigning different sizes to the simulated objects and a certain albedo to obtain their visual magnitudes, thus determining observable objects up to a certain visual magnitude. Sizes are determined by a cumulative size distribution function that follows a power law:

\begin{equation}\label{eq}
N(r) = N_0\bigg(\frac{R_0}{r}\bigg)^{\gamma},
\end{equation}

\noindent where $N$ is the number of objects with radius greater than $r$, $N_0$ is a scaling factor, $R_0$ is the radius above which there is just one object and $\gamma$ is an empirical parameter obtained from observations. We consider a size distribution with 3 slopes\footnote{In fact, OBIP never reaches the slope associated to the smallest sizes since only brighter objects are needed to be considered.} (\cite{Fraser2010}, \cite{Shankman2013}, \cite{Fraser2014}) $\gamma=3$, $\gamma=5$ and $\gamma=2$, respectively for radii in the range $r>200$ km, 50 km $<r<$ 200 km and $r<50$ km. The largest $r$ ($R_0$) is determined by associating it with the visual magnitude of 20.5, which is approximately the magnitude of the brightest TNO with 100 au $<a<$ 200 au and $q>30$ au. Albedos are taken as $0.25$.

This collection of sizes is randomly associated to the simulated particles' orbits which for the last $10^8$ yr belong to the range $100$ au $<a<$ $200$ au and $q>30$ au. The mean longitudes are chosen randomly in the range $0^{\circ}$ - $360^{\circ}$. Since there are $34$ real objects within the limits above, we determine the $34$ brightest objects by OBIP and the number of them with $q>42$ au. Another parameter that is considered by OBIP is the range of ecliptic latitude in which it will be applied, we consider a reasonable range of $\pm30^\circ$.

\begin{figure}[ht!]
\epsscale{1.1}
\plotone{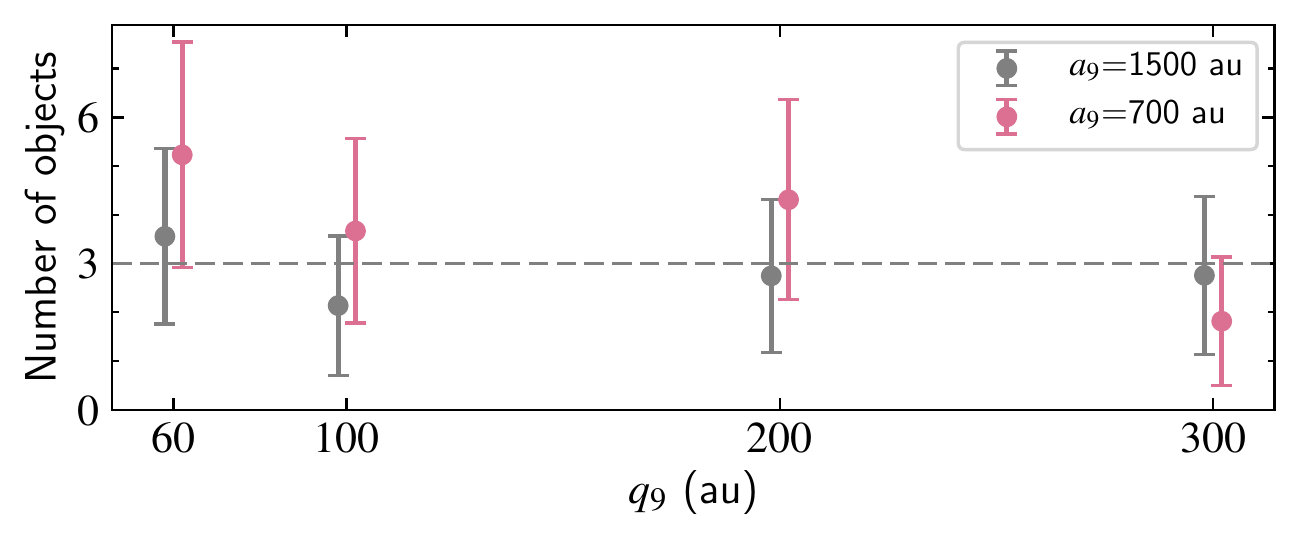}
\caption{Average of the number of observable objects by $1000$ realizations of OBIP, with $q>42$ au among the $34$ brightest ones with $q>30$ au and 100 au $<a<$ 200 au, for models with $i_9=30^\circ$ and $\omega_9=0^\circ$ as a function of the perihelion distance of Planet 9. The bars represent standard deviations of the mean values.\label{q9_vs_mean_ngt42}}
\end{figure}

The mean and standard deviation of the number of objects with $q>42$ au and $100$ au $<a<$ $200$ au among the $34$ brightest ones determined by $1000$ realizations of OBIP are depicted for each model in Table \ref{table3}. Although we can see some trend for a larger number of OBIP-determined objects with $q>42$ au for Planet 9 with a smaller $q_9$, this trend is not well defined as to place a low perihelion Planet 9 as prohibitive or even unlikely (see also Fig. \ref{q9_vs_mean_ngt42}). A typical example is Planet 9 with $a_9=1500$ au and $q_9=60$ au. OBIP determines an average of $3.56$ objects with $q>42$ au and standard deviation at $1.80$, thus $3$ objects is well within the range of statistically acceptable numbers. Most of Planet 9 models yield statistically coherent results with respect to the ratio of observable detached to scattering objects, what suggests that orbital parameters of Planet 9 with low $q_9$ should not be discarded as suggested in \citet{Brown2016}.

\begin{table}[h!]
\centering
\caption{Mean and standard deviation of the number of  objects with $q>42$ au among the $34$ brightest ones with $q > 30$ au, in the range 100 au $<a<$ 200 au.} \label{table3}
\setlength\tabcolsep{12pt} 
\begin{tabular}{l@{}>{\hspace*{1.2cm}} c c}
\tablewidth{0pt}
\hline
\hline
\multirow{2}{*}{Model} & \multicolumn{2}{c}{Number of objects with $q>42$ au} \\ \cmidrule(lr){2-3}
& Mean & Std \\
\hline
\decimals
$a$1500-10-0-100-5 & 5.69 & 2.25 \\
$a$1500-30-0-60-5 & 3.56 & 1.80 \\
$a$1500-30-0-100-5 & 2.14 & 1.43 \\
$a$1500-30-0-200-5 & 2.75 & 1.57 \\
$a$1500-30-0-300-5 & 2.76 & 1.62 \\
$a$1500-30-90-60-5 & 5.30 & 2.13 \\
$a$1500-60-0-100-5 & 5.79 & 2.18 \\
 $a$700-10-0-100-3 & 9.69 & 2.93 \\
  $a$700-30-0-60-3 & 5.23 & 2.31 \\
 $a$700-30-0-100-3 & 3.67 & 1.89 \\
 $a$700-30-0-200-3 & 4.31 & 2.05 \\
 $a$700-30-0-300-3 & 1.82 & 1.32 \\
 $a$700-30-90-60-3 & 3.35 & 1.77 \\
 $a$700-60-0-100-3 & 8.04 & 2.59 \\
\hline
\end{tabular}
\end{table}

\subsection{The influence of Planet 9 on the Classical Kuiper Belt}

\subsubsection{Cold Classical Kuiper Belt}

In order to analyze the influence of Planet 9 on the Cold Classical Kuiper Belt Objects (CCKBOs), we performed new numerical integrations for the models $a$700-30-0-60-3 and $a$1500-30-0-60-5. Both simulations were done with an integration time of 4.5 Gyr and initially considering the current orbital configuration of the four giant planets, the respective external perturber and a disk composed of 2000 test particles with 42.5 au $<a<$ 46 au, $0<e<0.01$, $0^\circ<i<0.6^\circ$ and remaining orbital elements randomly distributed between $0^\circ$ and $360^\circ$. We chose a conspicuously cold disk in order to check not only the compatibility of Planet 9 with the Cold Classical Kuiper Belt (CCKB) but also whether Planet 9 could be responsible for some of the excitation we currently find there.

\begin{figure}[ht!]
\epsscale{1.1}
\plotone{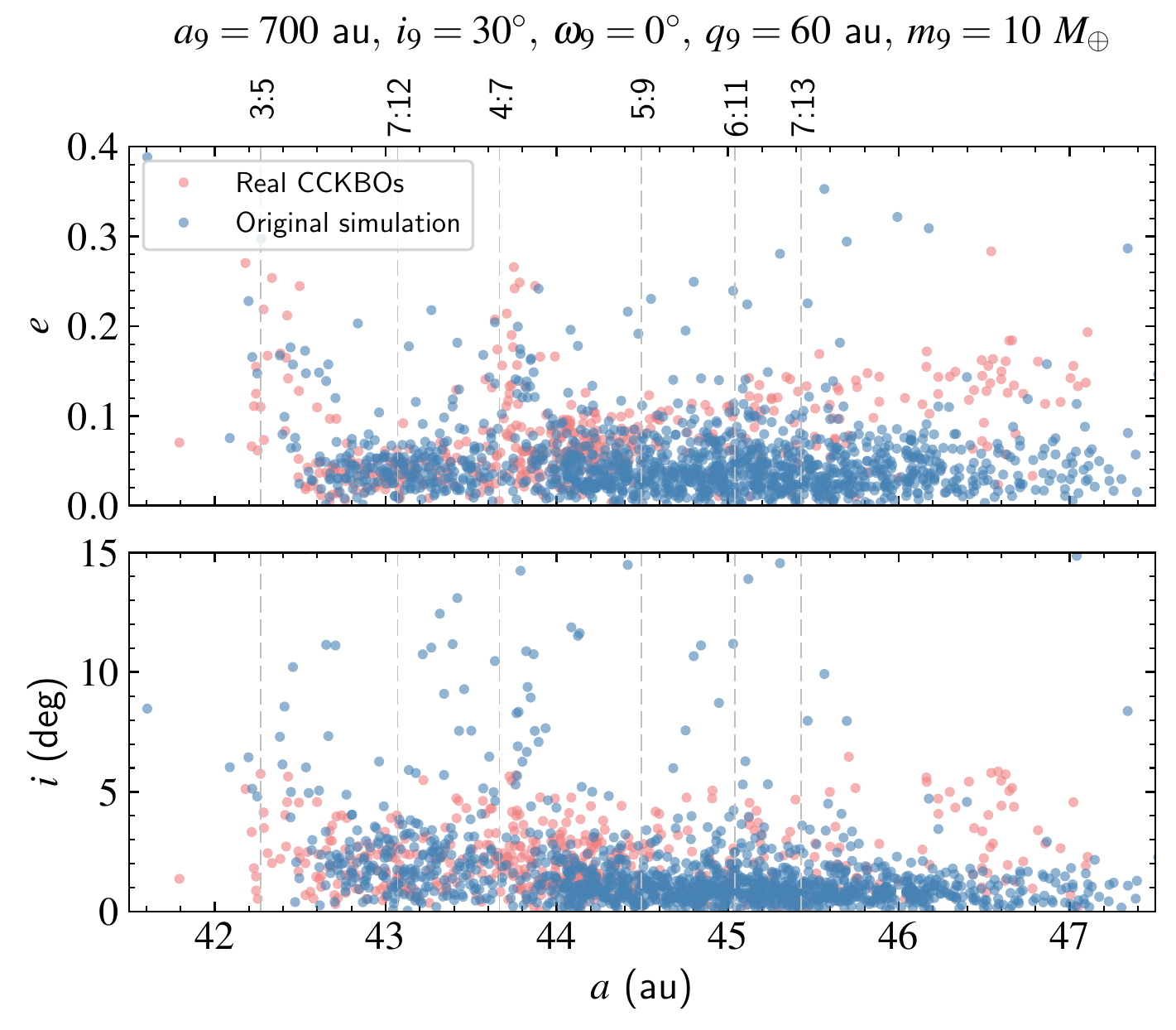}
\caption{Eccentricity and inclination as a function of the semimajor axis for real CCKBOs and simulated particles at 4.5 Gyr considering the model $a$700-30-0-60-3, in red and blue respectively. The simulated particles had initially semimajor axes between 42.5 and 46 au in almost circular and low inclination orbits. \label{KB700_obs_sim6_borda_46_0}}
\end{figure}

\begin{figure}[ht!]
\epsscale{1.1}
\plotone{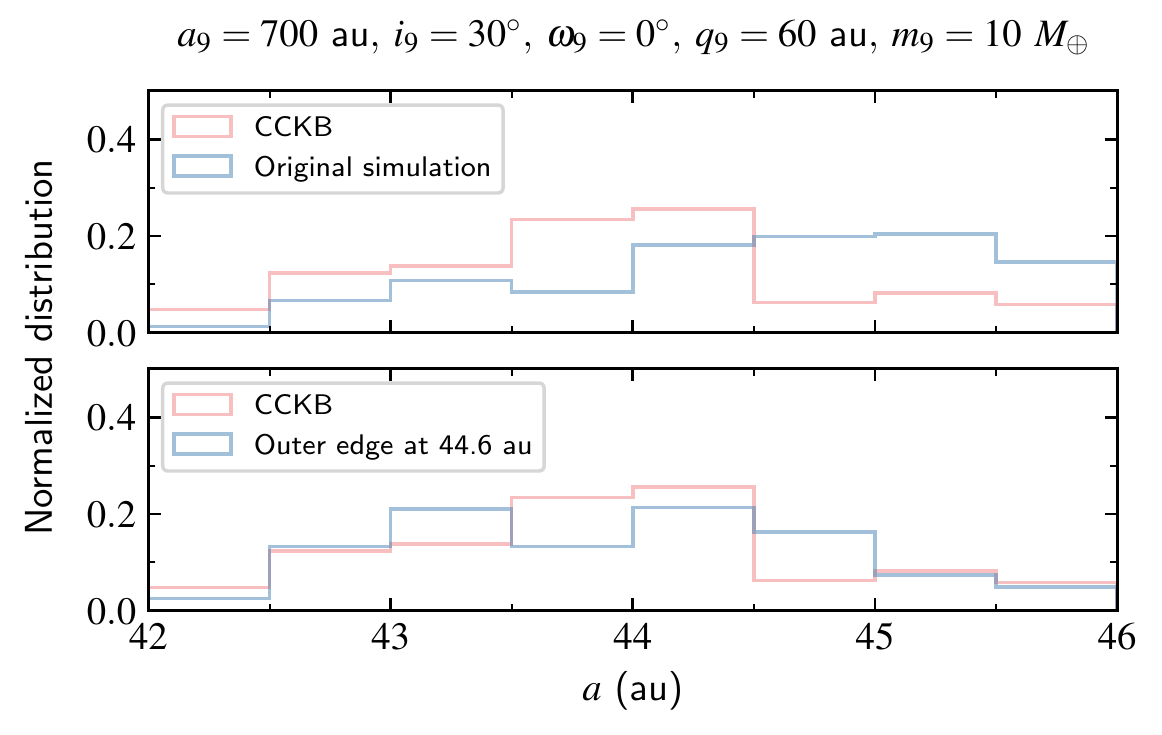}
\caption{Distribution of semimajor axes for actual CCKBOs and simulated particles at 4.5 Gyr for model $a$700-30-0-60-3, in red and blue respectively. Top panel shows the results for the full set of particles and the bottom one for the best fit in the truncation of the initial outer edge of the disk, at 44.6 au.\label{a700-30-0-60-3-histogramas}}
\end{figure}

\begin{figure}[ht!]
\epsscale{1.1}
\plotone{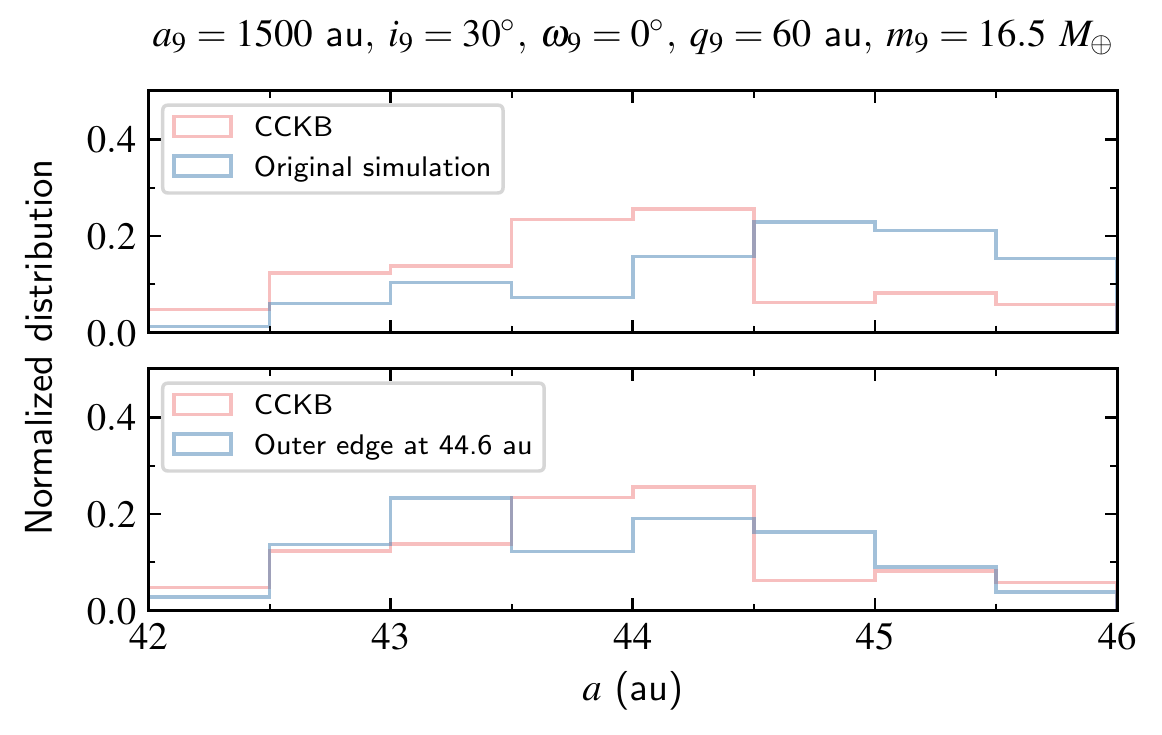}
\caption{Same as Fig. \ref{a700-30-0-60-3-histogramas} for the model a1500-30-0-60-5, where the best fit of the outer edge of the disk is at 44.6 au.\label{a1500-30-0-60-5-histogramas}}
\end{figure}

Both tested models present similar results (see e.g. Fig. \ref{KB700_obs_sim6_borda_46_0}). Planet 9 leads to the scattering of particles both inwards and outwards, although mostly outwards. They are however destabilized once they enter the secular resonance region, therefore, the main effect is to drive the semimajor axes of the particles to higher values. The eccentricities as well as the inclinations were excited within the values of the observed Cold population. Assuming a local formation, the original external border of the Cold disk is unknown. Thus, from the data of the original simulations we can obtain Cold disks with different external edges by restricting the initial semimajor axes of the particles; the best fit is then determined according to the final distribution of the semimajor axes of the particles in the interval 42 au $<a<$ 46 au, when compared to that of the actual Cold population. Figures \ref{a700-30-0-60-3-histogramas} and \ref{a1500-30-0-60-5-histogramas} show, for each model, the histograms in semimajor axis at 4.5 Gyr for the original disk and that corresponding to the best fit. These figures suggest that the fuzziness of the observed outer border of the CCKB could be a signature of the perturbation of a low perihelion Planet 9.

The eccentricities of the simulated particles have always low values, so they do not compare very nicely with the real eccentricity distribution mainly for higher semimajor axes ($a > 44.5$ au). Thus the peculiar distribution of the eccentricities of the CCKB would not be explained by the perturbations of such a planet with low perihelion. Some other mechanism must have acted to excite the eccentricities of these particles (e.g. a temporary excited Neptune in a planetary instability migration model \citep{Gomes2018}). The inclinations, on the other hand, are kept within the range of the observed inclinations in the CCKB.

Possibly the only inconsistency produced by a low perihelion Planet 9 on the CCKB refers to a dearth of simulated particles just past the 4:7 resonance. This is likely produced by the outward scattering of particles by Planet 9 and the barrier effect of the 4:7 resonance that must capture some outward migrating particles. 

\subsubsection{Hot Classical Kuiper Belt}

\begin{figure}[ht!]
\epsscale{1.1}
\plotone{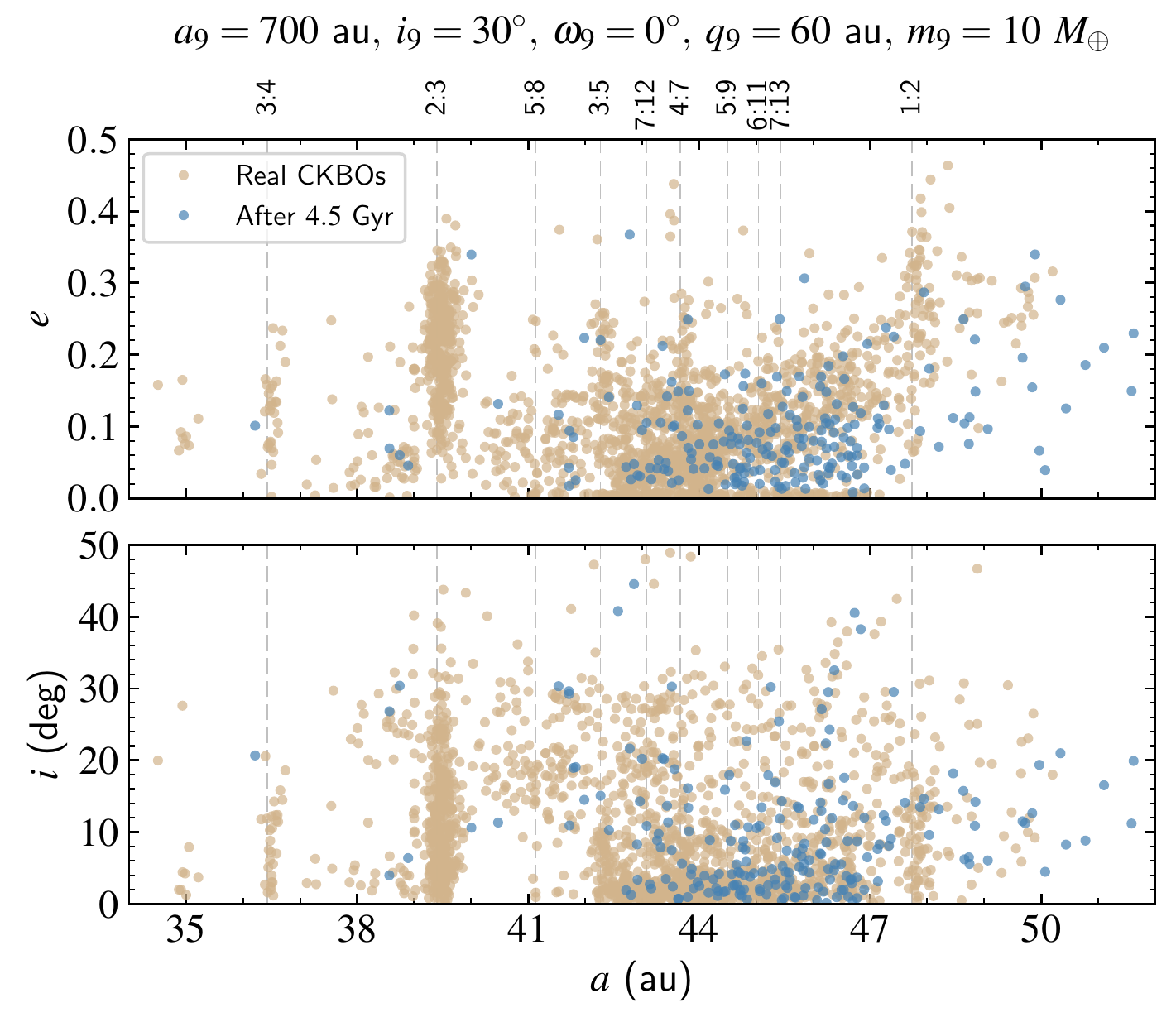}
\caption{Eccentricity and inclination as a function of the semimajor axis for actual CKBOs (in brown) and their configuration after 4.5 Gyr under the influence of Planet 9 under the model $a$700-30-0-60-3 (in blue). \label{a700-30-0-60-3-KB-real}}
\end{figure}

We also analyze the influence of Planet 9 on the Hot Classical Kuiper Belt objects. For this we performed simulations considering the actual Classical population (and resonant objects) of the Kuiper Belt for the model $a$700-30-0-60-3 and evolve the system by 4.5 Gyr. The orbital elements of the Kuiper Belt and the 4 known giant planets were taken for a same date from the JPL Small-Body Database\footnote{\url{https://ssd.jpl.nasa.gov/sbdb.cgi}}. The results show that a planet with such low perihelion would excessively clean objects with higher eccentricities (Fig. \ref{a700-30-0-60-3-KB-real}). These results led us to consider perihelion distances a little higher for the Planet 9.

\begin{figure}[ht!]
\epsscale{1.1}
\plotone{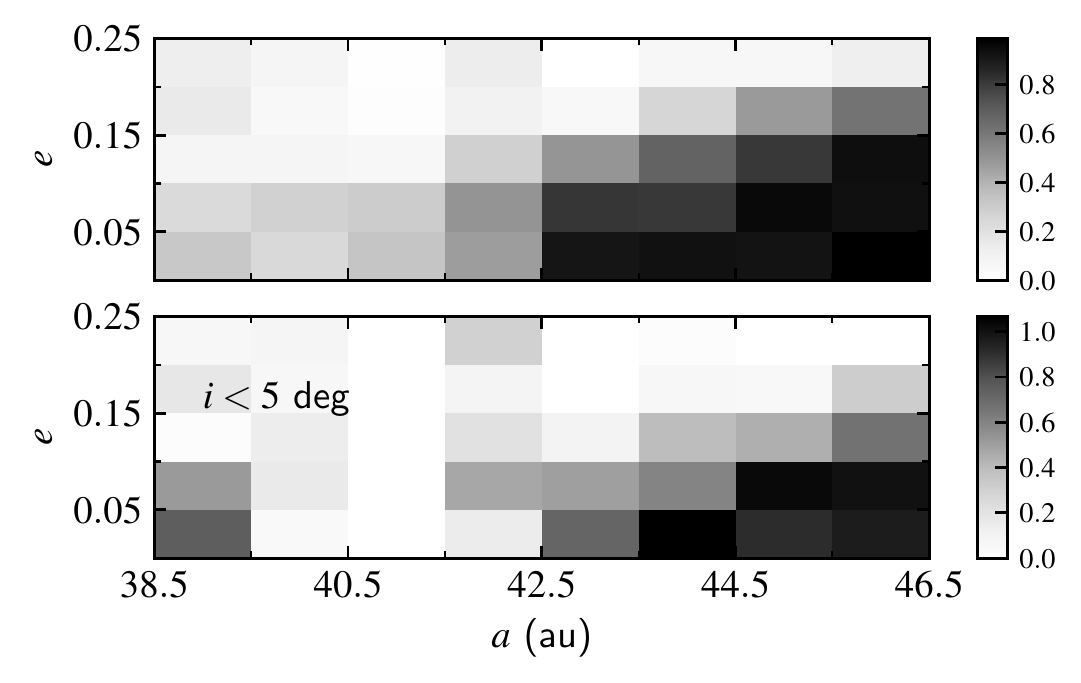}
\caption{Ratio of the number of surviving objects at the end of the integration time to the number of initial objects as function of the semimajor axis and eccentricity from an initially uniform distribution under the influence of the known planets. The upper figure considers all inclinations and the lower one only Cold objects ($i<5^{\circ}$).\label{sempl9-axe}}
\end{figure}

\begin{figure}[ht!]
\epsscale{1.1}
\plotone{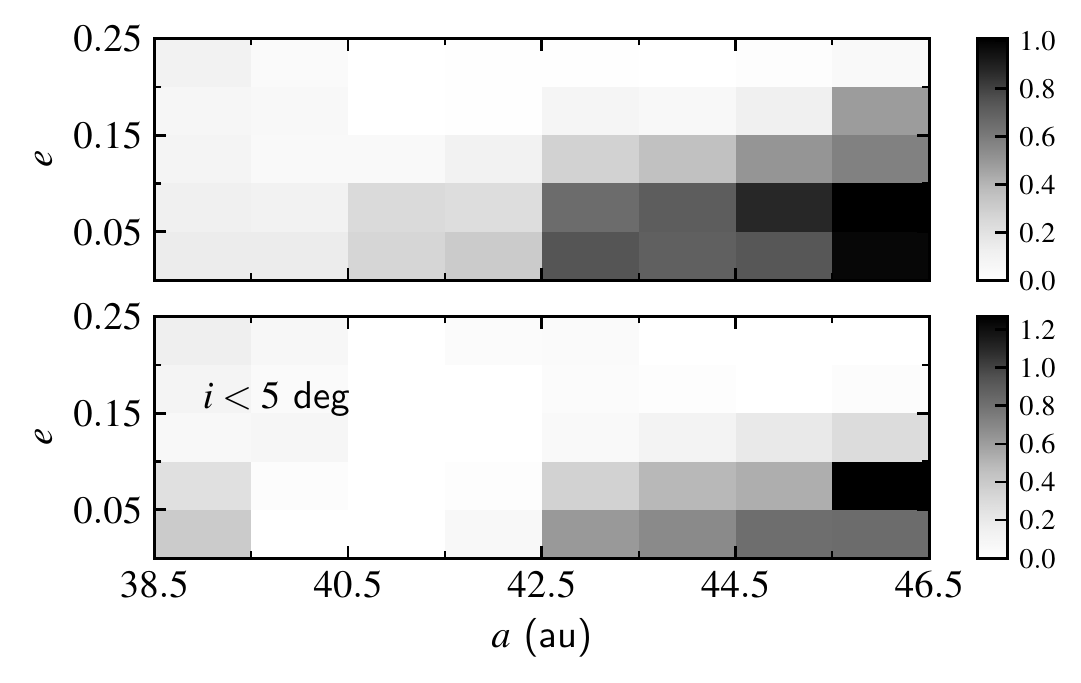}
\caption{Same as Fig. \ref{sempl9-axe} adding a planet with $a_9=700$ au and $q_9=70$ au.\label{700-0_9-axe}}
\end{figure}

\begin{figure}[ht!]
\epsscale{1.1}
\plotone{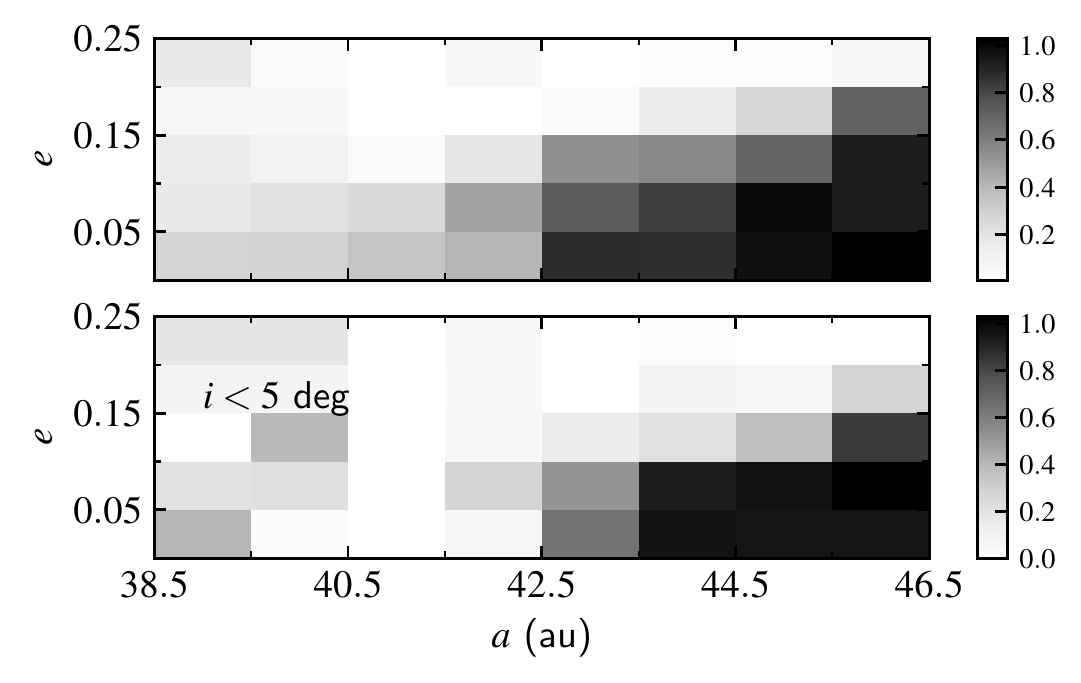}
\caption{Same as Fig. \ref{sempl9-axe} adding a planet with $a_9=1500$ au and $q_9=90$ au.\label{1500-0_94-axe}}
\end{figure}

We thus performed numerical integrations considering a theoretical distribution of particles uniformly and randomly distributed in the ranges 38.5 au $<a<$ 46.5 au, $0<e<0.25$ and $0^\circ<i<20^\circ$ under the influence of the known giant planets and compared the results with two other models that additionally consider a Planet 9 with a low perihelion but larger than 60 au, namely, $a_9=700$ au and $q_9=70$ au and, $a_9=1500$ au and $q_9=90$ au. The results at 4.5 Gyr are presented in Figs. \ref{sempl9-axe}, \ref{700-0_9-axe} and \ref{1500-0_94-axe} as the ratio of the number of surviving objects to the number of initial objects in the grid of the plane $a-e$ for objects with any inclination and for objects with $i<5^\circ$. 

Figure \ref{sempl9-axe} shows that a model without an additional external planet tends to preserve more hot and cold objects at the lower right end of the plane $a-e$. This is already well known since this is the region that keeps objects far from close encounters with Neptune. Moreover, the region between roughly $40$ and $42$ au is also depleted due to the $\nu_8$ secular resonance.

Now comparing Figs. \ref{sempl9-axe}, \ref{700-0_9-axe} and \ref{1500-0_94-axe} we notice that the model with an external perturber with $a_9=1500$ au and $q_9=90$ au preserves the characteristics of the theoretical Kuiper Belt at 4.5 Gyr seen in the model considering only the known giant planets. The hot objects show a good approximation and most objects in the 3:5 and 4:7 MMRs (corresponding to $a=42.3$ au and $a=43.7$ au) survived until the end of the integration. The model that considers a planet with $a_9=700$ au and $q_9=70$ au has a smaller fraction of hot objects compared to the previous one and does not keep most objects in MMR. The right comparison of the distribution of TNOs in the Kuiper Belt from simulations with the observed ones makes sense only if the right initial conditions are considered. For instance, the initial distribution of objects after the stabilization of the giant planets\footnote{In a Nice model scenario, but also a smoother migration scenario could work in this case.} might have concentrated most objects in smaller semimajor axes. Planet 9 could thus shift the semimajor axes outwards towards a distribution similar to our present one. Apart from this speculation, these last results anyway show that a Planet 9 with a perihelion distance as low as 90 au does not significantly affect the Classical Kuiper Belt.

\section{Conclusions}\label{conclusions}

An external perturber was proposed to explain the confinements in the orbital elements $\Omega$, $\omega$ and $\varpi$ of TNOs with $q>30$ au and $a>250$ au. \cite{Brown2016} claimed that good orbits for Planet 9 should have $q_9 > 200$ au, to explain such confinements, where lower perihelia than that above were assigned a null probability arguing that these parameters produce objects with $q>42$ au in the region 100 au $<a<$ 200 au, where no object with such perihelion distance had been observed previously. On the other hand, \cite{Gomes2017} explain the inclination of the solar equator in relation to the invariable plane of known planets showing that parameters of the Planet 9 similar to those found by \citet{Batygin2016} and \citet{Brown2016} are compatible although with slightly higher eccentricities.

In this way, we considered perihelion distances for Planet 9 lower than those indicated by \citet{Brown2016} and we tested their influence on the TNOs with $a\geq250$ au and $q\geq40$ au, since these objects would not be greatly influenced by Neptune. In this region there are currently 9 objects whose $\Omega$ exhibit a large confinement ($210.5^\circ$), while the confinement in $\omega$ is $131.7^\circ$ and we assume two confinements in the observational sample of $\varpi$.

Through statistical analysis we tested the influence of a Planet 9 with $a_9=700$ au and $a_9=1500$ au and perihelion distances ranging from 60 to 300 au. Our results show that smaller perihelion distances ($q_9=60$ au or $q_9=100$ au) produce narrower confinements. Such confinements are generally best reproduced in $\varpi$ and interestingly in several models there are two confinements in this angular element, as assumed in the observational data, being shepherded by Planet 9. Models with high $q_9$ are generally associated with random distributions in $\Omega$ and $\omega$. 

We also verified that a planet with low perihelion roughly yielded the current observed ratio between the number of scattered and detached with $q>42$ au objects in the interval 100 au $<a<$ 200 au by applying an observational bias to the particles of our models. We saw that several of our models are compatible with this ratio. Likewise, we verified whether the existence of a Planet 9 with such low perihelion would destroy the Classical Kuiper Belt. Integrations considering the most extreme cases of perihelion distance of Planet 9 from our models on an initially cold disk show that the signature due to the external perturber would be that of moving the semimajor axes of the objects to larger values and the sparseness of particles immediately past the 4:7 MMR. However, when we considered also the hot objects, a Planet 9 with a perihelion at $60$ au would considerably deplete the region with a particularly important depletion for the resonant objects.

This motivated us to perform other runs with low perihelion distances but larger than 60 au. For this we considered a theoretical Classical Kuiper Belt and showed that a Planet 9 with $q_9=90$ au produces signatures similar to those produced without an additional external planet. Thus, we concluded that planets with perihelion distances as small as $90$ au  should not be discarded in principle. They produce better confinements while preserving the Classical Kuiper Belt and the ratio between the number of detached to scattering objects in the region 100 au $<a<$ 200 au.

A last remark is that although a very low perihelion distance for Planet 9 ($\sim 60 - 70$ au) seems to excessively empty the Kuiper Belt region, this may not be too bad, depending on the initial conditions of the particles just after the stabilization of the planets. If these objects were much more concentrated on smaller semimajor axes, a low perihelion Planet 9 could drift these objects outward so as to populate the Kuiper Belt conveniently. This problem is the subject of a following work by the authors. 

\acknowledgments

The numerical simulations were done by the Lobo Carneiro supercomputer in the High Performance Computing Center of NACAD/COPPE/UFRJ. J.C. thanks support from the Coordination for the Improvement of Higher Education Personnel (CAPES) and R.G. acknowledges the Brazilian CNPq grant through process 307009/2014-9. 

\bibliographystyle{aasjournal}
\bibliography{Bibliography/bibliography}

\end{document}